\documentstyle[prl,aps,amsfonts,amssymb,twocolumn,epsfig]{revtex}
\begin{document}
\draft

\newcommand{\w}{\omega}
\newcommand{\kf}{k_F}
\newcommand{\Q}{{\bf Q}}
\newcommand{\A}{\text{\AA}}
\newcommand{\IM}{\text{Im}}
\renewcommand{\vec}[1]{{\bf #1}}
\wideabs{
\title{Interplay of disorder and 
spin fluctuations in the resistivity near a quantum critical point}
\author{A. Rosch}
\address{Serin Laboratory, Rutgers University, Piscataway, NJ
08854-0849, USA}
\date{\today}
\maketitle

\begin{abstract}
  The resistivity in metals near an antiferromagnetic quantum critical
  point (QCP) is strongly affected by small amounts of disorder. In a
  quasi-classical treatment, we show that an interplay of strongly
  anisotropic scattering due to spin fluctuations and isotropic
  impurity scattering leads to a large regime where the resistivity
  varies as $T^\alpha$, with an anomalous exponent, $\alpha$,
  $1\lesssim \alpha \lesssim 1.5$, depending on the amount of
  disorder.  I argue that this mechanism explains in some detail the
  anomalous temperature dependence of the resistivity observed in
  CePd$_2$Si$_2$, CeNi$_2$Ge$_2$ and CeIn$_3$ near the QCP.
\end{abstract}
}

In the last five years an increasing number of heavy-fermion metals
near an antiferromagnetic (AFM) quantum critical point was shown to
display striking deviations from conventional Fermi-liquid behavior
\cite{loehneysen,julian,mathur,gegenwart}.  A few of the dirtier systems
(e.g. \cite{gegenwart}) appeared to show the $T^{3/2}$ behavior of the
resistivity, $\rho (T)$, as predicted by the Hertz-Millis
spin-fluctuation theory for such a QCP \cite{hertz,millis,moriya}. In
other cleaner single-crystal systems, such as CePd$_2$Si$_2$,
CeNi$_2$Ge$_2$ and CeIn$_3$ \cite{julian,mathur}, $\rho(T)$ varies as
$T^{\alpha}$ with exponents $\alpha$ between $1.1$ and $1.5$.  It has
been argued \cite{coleman} that such a behavior signals a fundamental
breakdown of Fermi liquid theory.

In this letter we propose a simple theoretical explanation for this
anomalous behavior which covers both dirty and clean systems.  Our
main result is that the resistivity anomalies mentioned above can be
attributed to the interplay between quantum-critical AFM spin
fluctuations~\cite{hertz,millis} and impurity scattering in a
conventional Fermi liquid.  In fact, the resistivity can be described
in semi-quantitative terms in the context of the simplest
semi-classical Boltzmann equation approach.  Here, we will have
nothing to say about the striking linear behavior of $\rho(T)$ near the
QCP in CeCu$_{6-x}$Au$_x$ \cite{loehneysen} in which the neutron scattering
experiments suggest a more unconventional behavior of the
spin-fluctuations \cite{rosch}.

\begin{figure}[t]
  \centering
 \epsfig{width=.8 \linewidth,file=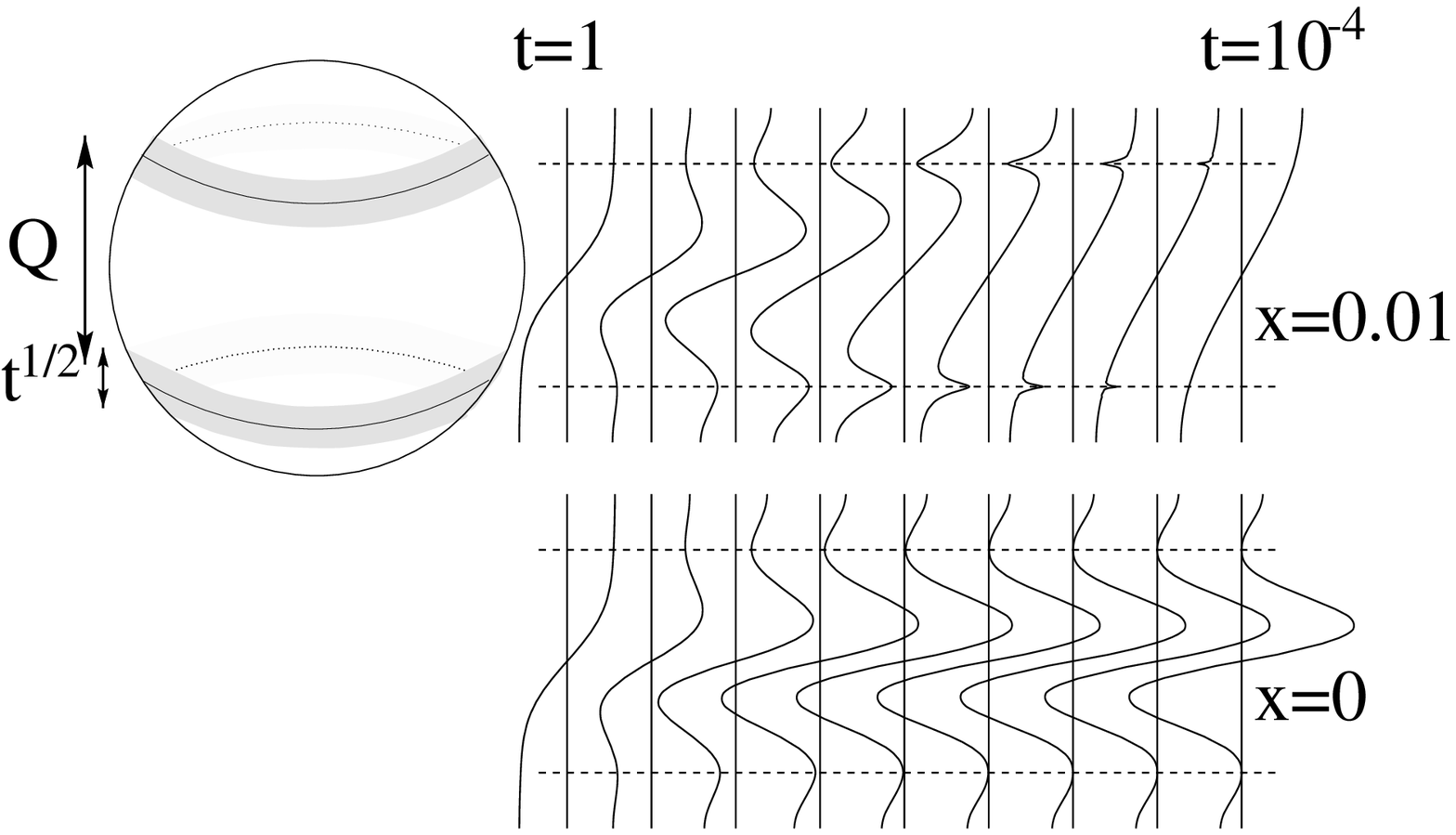}
\caption[]{Near the transition to an antiferromagnet with ordering vector 
${\vec{Q}}$, the scattering on the 
Fermi-surface is enhanced along ``hot lines'' connected by ${\vec{Q}}$.
The width of this region is given by $\Delta k \approx q_0 \sqrt{T/\Gamma}$.
The out-of-equilibrium distribution $\Phi_{\theta}$ of the quasi particles
for a current parallel to $\vec{Q}$ is shown as a function of the azimuthal
angle $\theta$ for temperatures ranging
from $t=(q_0/\kf)^2 (T/\Gamma)^2=1$ (left) down to $t=10^{-4}$ (right)
and for both a clean system (lower curves) 
and  a small amount of disorder,
$x=0.01$ (upper curves).
}
\label{figHot}
\end{figure}

The temperature dependence of $\rho(T)$ can already be understood from
the following qualitative argument: First we recall that scattering
off AFM spin-fluctuations is most effective near ``hot lines'', i.e.,
points at the Fermi surface connected by the ordering wave vector $\Q$
(see Fig.~\ref{figHot}) where gaps would open up in the
antiferromagnetically ordered (metallic) phase.  As explained by
Hlubina and Rice \cite{hlubina}, the strong scattering near these
lines is short circuited by the ``cold" regions on the Fermi surface,
where the scattering rates are small. In clean systems the latter
dominate the transport and resistivity acquires the usual Fermi liquid
behavior, $\rho(T) \sim T^2$ at sufficiently low temperatures. 
Impurity scattering
leads essentially to an averaging of the scattering rate over the
Fermi surface reducing the effectiveness of the Hlubina-Rice mechanism
and emphasizing the role of the ``hot lines".  The above line of
argumentation can be made more quantitative: Near the ``hot lines''
the scattering rate $1/\tau_S$ of the quasi-particles due to the
quantum-critical spin-fluctuations is linear in temperature
$\tau_M/\tau_S \approx t=T/\Gamma$ where $\Gamma$ is a characteristic
energy scale and $\tau_M$ a typical scattering time. The width of
these ``hot lines'' is given by $\sqrt{t}$ (see below). In the ``cold
regions'', we expect Fermi liquid behavior with $\tau_M/\tau_S \approx
t^2$. Weak disorder leads to an isotropic scattering rate,
$1/\tau_{\text{el}}=x/\tau_M$, where $x$ is a small dimensionless
number measuring the effectiveness of impurity compared to magnetic
scattering ($x^{-1}\approx k_F l$ for spin-fluctuation scattering in
the strong coupling ``unitarity" limit; $l$ is the elastic
mean-free-path).  The conductivity $\sigma$ is proportional to the
average of $\tau_{\vec{k}}$ over the Fermi surface (FS) with
$1/\tau_{\vec{k}}=1/\tau_{\text{el}}+1/\tau_S(\vec{k})$:
\begin{eqnarray}
  \label{qualitative}
  \sigma \propto \left< \tau_{\vec{k}} \right>_{\text{FS}}
\propto \frac{\sqrt{t}}{x+t}+\frac{1-\sqrt{t}}{x+t^2}.
\end{eqnarray}
The two terms describe the contributions from ``hot" and ``cold" 
regions, respectively.
For $x<t^2<1$ cold regions short-circuit
the hot ones and from the second term in (\ref{qualitative}) we obtain
$\rho \propto t^2$  \cite{hlubina};
while for 
$t<x<1$, the hot spots dominate and the resistivity $1/\sigma$ is
proportional to $x+t^{3/2}$.  
At intermediate temperatures, $x<t<\sqrt{x}$, we expect a crossover
regime, in which we can define an effective resistivity exponent $\alpha$,
in terms of the 
logarithmic difference of $\Delta
\rho(T)=\rho(T)-\rho(0)$ at the crossover temperatures $T^c_1=\Gamma
x$ and $T^c_2=\Gamma \sqrt{x}$,
\begin{eqnarray}
\alpha \approx \frac{\ln \Delta \rho(T^c_2)-\ln \Delta \rho(T^c_1)}{
\ln T^c_2-\ln T^c_1} \approx \frac{\ln x-\ln x^{3/2}}{
\ln \sqrt{x}-\ln x}=1.\label{alpha}
\end{eqnarray}
In a very clean system this crude estimate implies
a nearly linear crossover behavior in temperature as measured in 
clean samples of CePd$_2$Si$_2$ and CeNi$_2$Ge$_2$ \cite{julian,mathur}.

We now proceed with a more precise argument based on the
semi-classical Boltzmann equation treatment of
electrons interacting with spin-fluctuations and impurities. 
The former are described
by a theory above the upper critical dimension \cite{hertz,millis} and,
due to the ohmic damping of
the magnetic excitations in a metal, is
characterized by a dynamical exponent $z=2$.
As a result, the spin-fluctuation spectrum can be modeled by
\cite{hertz,millis}
\begin{eqnarray} \label{chi}
\chi_{\vec{q}}(\w)=\chi_{-\vec{q}}(\w)\approx 
\frac{1}{1/(q_0 \xi)^2+(\vec{q}\pm \vec{Q})^2/q_0^2- i \w/\Gamma},
\end{eqnarray}
where $q_0$ and $\Gamma$ are characteristic momentum and energy scales;
and $\xi$ is the AFM correlation length which, at the QCP,
diverges as $1/\xi^2 \approx c q_0^2 (T/\Gamma)^{3/2}$
\cite{hertz,millis}.
For the purposes of our numerical calculations we set $c=1$ (
$c$ does not influence the low-temperature properties).

Our starting point is the Boltzmann equation with a quasi-particle
distribution function
$f_{\vec{k}}=f^0_{\vec{k}}-\Phi_{\vec{k}} (\partial
f^0_{\vec{k}}/\partial \epsilon_{\vec{k}})$ linearized around
the Fermi distribution $f^0_{\vec{k}}$ with a  collision term
\begin{eqnarray} 
\left.\frac{ \partial f_{\vec{k}}}{\partial t}\right|_{\text{coll}} \!\!\!
&=&  
\sum_{\vec{k}'} \frac{f^0_{\vec{k}'} (1-f^0_{\vec{k}})}{T} 
(\Phi_{\vec{k}}-\Phi_{\vec{k}'})\nonumber \\
&& \hspace{-1.5cm} \times
\left[ 
g_{\text{imp}}^2 
\delta(\epsilon_{\vec{k}}-\epsilon_{\vec{k}'})+\frac{2 g_S^2}{\Gamma}
n^0_{\epsilon_{\vec{k}}\!-\!\epsilon_{\vec{k}`}}
\IM \chi_{\vec{k}-\vec{k}'}(\epsilon_{\vec{k}}\!-\!\epsilon_{\vec{k}`})
\right] \label{coll}
\end{eqnarray}
Here $g_{\text{imp}}^2$ and $g_S^2$ are transition rates for isotropic
impurity scattering and inelastic scattering from spin fluctuations,
respectively, and $n^0_{\w}$ is the Bose function.  Eqn. (\ref{coll})
tacitly assumes that the spin-fluctuations stay in equilibrium, an
approximation valid if the spin fluctuations can loose their momentum
effectively
by Umklapp or impurity scattering.
Instead of solving the Boltzmann equation directly, it is instructive
to consider a mathematically equivalent variational
principle \cite{ziman}.  Following Hlubina and
Rice \cite{hlubina}, we define a functional $\rho$ of $\Phi_{\vec{k}}$
\begin{eqnarray}
\rho[\Phi_{\vec{k}}]&=& \frac{\hbar}{4 e^2}\frac{\oint\!\!
 \oint
\frac{d \vec{k}d \vec{k}`}{v_{\vec{k}}v_{\vec{k}'}}
 F_{\vec{k}\vec{k}'} (\Phi_{\vec{k}}-\Phi_{\vec{k}`})^2}
{\left(\oint\frac{d \vec{k}}{v_{\vec{k}}} (\vec{v}_{\vec{k}} \vec{n})
\Phi_{\vec{k}} \right)^2} 
\,\, \longrightarrow \,\,\text{min} \label{min1}\\
F_{\vec{k}\vec{k}`}&=&g_{\text{imp}}^2+\frac{2 g_S^2}{\Gamma T} 
\int_0^{\infty}
\w n^0_{\w} [ n^0_{\w}+1] \IM \chi_{\vec{k}-\vec{k}`}(\w) d \w. \label{F}
\end{eqnarray}
The physical resistivity is given by the minimum of
$\rho[\Phi_{\vec{k}}]$ regarded as 
a functional of  $\Phi_{\vec{k}}$.
At this minimum $\Phi_{\vec{k}}$ is directly proportional to
the true distribution
function in an electric field applied in the direction
of the unit vector $\vec{n}$. In the above expression only integrals
over the Fermi surface enter, as we have already integrated out
the perpendicular components of $\vec{k}$ 
by using $\int d \vec{k}=\int d \epsilon_{\vec{k}} \oint d
\vec{k}/v_{\vec{k}}$ where $v_{\vec{k}}$ is the velocity of quasi
particles at the Fermi surface.

In the following we simplify the discussion by (i) considering a
spherical Fermi surface and (ii) limiting ourselves to transport
parallel to the ordering wave vector $\vec{Q}=(0,0,2 \kf \cos
\theta_H)$, in which case $\Phi_{\vec{k}}=\Phi_{\theta}$ is only a
function of the azimuthal angle $\theta$.  The equations
$\theta=\theta_H$ and $ \theta=\pi-\theta_H$ define the ``hot lines''
shown in Fig.~\ref{figHot}. The geometry \cite{perp} and precise value
of $\theta_H$ are not very important as long as one stays away from
$\theta_H=0$ (``$2 \kf$'' ordering) or $\theta_H=\pi/2$ (ferromagnetic
ordering) where our approach breaks down \cite{millis}; in our
numerical calculations we use $\theta_H=\pi/6$. As in \cite{hlubina}
we approximate the second term in Eqn.~(\ref{F}) by $ 2 g_S^2 I\left[y
\right]$ with $I[y]\approx \pi^2/( y (3 y+ 2 \pi))$ and $y= (\Gamma/T)
\left(1/(q_0 \xi)^2+(\vec{q}\pm \vec{Q})^2/q_0^2 \right)$ which is
asymptotically exact for large and small $y$.

After performing the integration over $\w$ and  the polar angle $\varphi$
in Eqn. (\ref{min1})
we obtain
\begin{eqnarray}
  \label{rhoMin}
  \rho(T)&=& \text{min}\!\left[
          \rho_{\text{imp}}[\Phi_{\theta}]+\rho_{S}[\Phi_{\theta}] 
 \right]\\
 \rho_{\text{imp}}&=& \frac{x \rho_M}{6} 
\frac{\int\!\!\!\int_0^{\pi} 
(\Phi_{\theta_1}\!-\!\Phi_{\theta_2})^2 \sin \theta_1  d  \theta_1 
  \sin \theta_2 d  \theta_2}
{\left(\int_0^{\pi} \cos \theta \Phi_{\theta}
\sin \theta d \theta\right)^2} \label{rhoImp}\\
 \rho_{S}&=& \frac{\pi \rho_M }{3}
\frac{\int\!\!\!\int_0^{\pi} F_{\theta_1 \theta_2}
(\Phi_{\theta_1}\!-\!\Phi_{\theta_2})^2 \sin \theta_1  d  \theta_1 
  \sin \theta_2 d  \theta_2}
{\left(\int_0^{\pi} \cos \theta \Phi_{\theta}
\sin \theta d \theta\right)^2}  \label{rhoS}
\end{eqnarray}
where $\rho_M=3 \hbar g_{S}^2 /(e^2 v_F^2)$ is a typical resistivity
due to scattering from spin fluctuations at approximately the
temperature scale $\Gamma$ and $x= g_{\text{imp}}^2/(2 g_S^2)$
measures the relative strength of impurity scattering. The prefactors
are chosen in such a way that $x \rho_M=3 \hbar g_{\text{imp}}^2 /(2
e^2 v_F^2)$ is the residual resistivity.  The dimensionless scattering
rate $F_{\theta_1 \theta_2}$ averaged over the polar angle $\varphi$
is given by
\begin{eqnarray}
F_{\theta_1 \theta_2}& \approx& \int_0^{2 \pi} \frac{d\varphi}{2 \pi}
 I\left[y(\theta_1,\theta_2,\varphi) \right] \label{Ftheta}\\
&\approx&\frac{\pi t^2/(2 \sin \theta_H) }
{ |\Delta \vartheta| \left( 2 t+\frac{3}{\pi}
        \left((\Delta \vartheta)^2+|\Delta \vartheta| 
  \sqrt{(\Delta \vartheta)^2+ 
\frac{2 \pi t}{3}}\right)\right)}\nonumber\\
(\Delta \vartheta)^2&=&
\vartheta_1^2+\vartheta_2^2+2 \vartheta_1 \vartheta_2
\cos 2 \theta_H+1/(\kf \xi)^2\nonumber
\end{eqnarray}
$\vartheta_1=\theta_1-\theta_H$ and
$\vartheta_2=\theta_2-(\pi-\theta_H)$ measure the distances from the
``hot lines'' and $t=(T/\Gamma) (q_0/\kf)^2$ is the dimensionless
temperature.  Our numerical calculations use the full polar integral in
(\ref{Ftheta}), even though at low temperatures,
($1/(\xi \kf)^2, t \ll 1$) the correct behavior is also contained
in the approximate form.

From Eqns. (\ref{rhoMin})--(\ref{Ftheta}) one can easily deduce the
qualitative behavior of the low-temperature resistivity.  We first
consider the case of very low temperatures in a dirty metal.  In this
regime the resistivity is dominated by the disorder contribution
$\rho_{\text{imp}}$. From $\delta
\rho_{\text{imp}}[\Phi_{\theta}]/\delta\Phi_{\theta}=0$ we find the
usual quasi-particle distribution for impurity scattering
$\Phi_{\theta}=\cos \theta$.  The leading temperature dependent
correction to the residual resistivity $\rho_0= x \rho_M$ is then
given by $\rho_{S}[\cos \theta]$. The main contribution to $\rho_{S}$
arises in a small region around the hot spots, where $
(\Phi_{\theta_1}-\Phi_{\theta_2})^2 \approx (\cos \theta_H-\cos (
\pi-\theta_H))^2$ is finite. Scaling $\vartheta_{1/2}$ in Eqn.
(\ref{Ftheta}) with $\sqrt{t}$ one recognizes that in the regime $
t>1/(\kf \xi)^2\propto t^{3/2}$ the finite correlation length can be 
neglected and at lowest temperatures the resistivity is given by
 \begin{eqnarray}
   \label{rhoDirty}
   \rho(T\to 0)=\rho_M\!\left[x + \sqrt{\frac{3 \pi^7}{8}}
\frac{q_0^3 \cos{\theta_H}}{\kf^3} 
\left(\frac{T}{\Gamma}\right)^{3/2}\right].
 \end{eqnarray}

\begin{figure}[t]
  \centering
 \epsfig{width=0.7 \linewidth,file=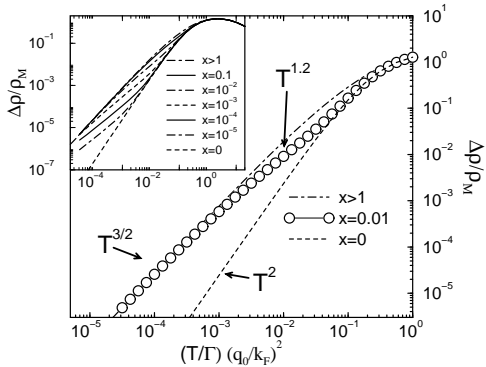}
\caption[]{Log-log plot
  of $\Delta \rho= \rho(T)-\rho(0)$ for a rather clean system with
  $x=0.01$.  Note the large crossover regime from the resistivity of a
  clean system (dashed line) at high temperatures to the resistivity
  of a dirty system (dot-dashed line). The inset shows, how this
  crossover evolves for various impurity concentrations $x$.}
\label{figRhoLogLog}
\end{figure}
On the other hand, if the system is clean ($x=0$), we have to minimize
$\rho_S[\Phi_{\theta}]$. As pointed out by Hlubina and Rice
\cite{hlubina}, the ansatz $\Phi_{\theta}=\cos \theta$ is far from the
true minimum. We can considerably reduce $\rho_S$ by using a
distribution function where the ``hot lines'' are excluded, e.g.
$\Phi_{\theta}=0$ for $|\theta-\theta_H|<\theta_{\text{cut}}$ and
$|\theta-(\pi-\theta_H)|<\theta_{\text{cut}}$. 
For this ansatz, the temperature
dependence in the numerator of the scattering rate (\ref{Ftheta}) can
be neglected for $\theta_{\text{cut}}\gg \sqrt{t}$ and the resistivity
is given by
\begin{eqnarray}
  \label{rhoClean}
  \rho(x=0,T\to 0)=c \rho_M (q_0/\kf)^2 (T/\Gamma)^2
\end{eqnarray}
where $c$ is a non-universal number of order $1$ which depends on the
details of the scattering mechanism in the ``cold
regions'' of the Fermi surface.

To obtain the crossover behavior we calculate the distribution function
$\Phi_{\theta}$ and the resistivity $\rho(T)$ within our model
numerically by solving
the integral equation $\delta
\rho[\Phi_{\theta}]/\delta\Phi_{\theta}=0$ which is equivalent to
solving the linearized Boltzmann equation directly. For a clean
system, $\Phi_{\theta}$ is shown in the lower part of
Fig.~\ref{figHot}.  At high temperatures the distribution function is
structureless and all parts of the Fermi surface (besides those
perpendicular to the current) contribute more or less equally to the
resistivity.  However, for lower temperatures the region around the
``hot lines'' (dashed lines in Fig.~\ref{figHot}) are
short circuited and the distribution function vanishes.  
Accordingly the resistivity is
much lower and drops $\propto T^2$ (dashed line in
Fig.~\ref{figRhoLogLog}).

\begin{figure}[t]
  \centering
 \epsfig{width=0.7 \linewidth,file=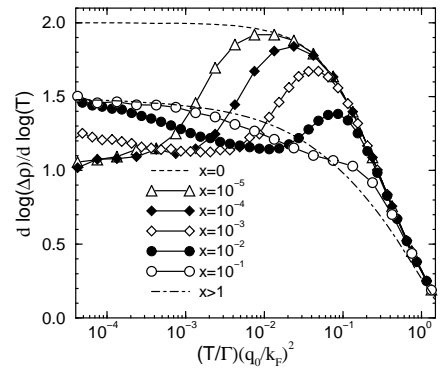}
\caption[]{Effective exponent of the resistivity, defined as the logarithmic
  derivative of $\Delta \rho(T)$. At very low temperatures, the
  ``dirty-limit'' exponent $3/2$ is recovered. However, in the
  experimentally accessible low temperature regime smaller exponents
  are to be expected for rather clean system ($x<0.1$).}
\label{figExponent}
\end{figure}
As shown in the simple calculation discussed in the beginning, in a
system with a small amount of disorder, we expect a large crossover
regime between the behavior described by Eqns.  (\ref{rhoDirty}) and
(\ref{rhoClean}). In the variational approach given here this is due
to the effect, that the impurity resistivity is {\em not} minimized by
the distribution function $\Phi_{\theta}^{\text{clean}}$ (the low
temperature curve in the lower part of Fig.~\ref{figHot}) and
$\rho_{\text{imp}}[\Phi_{\theta}^{\text{clean}}]= (1+c') x \rho_M$,
where $c'$ is a number of order $1$ (e.g. $c'\approx 2.8$ in our
model).  Below a temperature $T^c_2$, defined by $c' \rho(T=0)\approx
\rho(x=0,T^c_2)$, the distribution function deviates from
$\Phi_{\theta}^{\text{clean}}$ and approaches the $\cos \theta$-form
which minimizes impurity scattering (see Fig.~\ref{figHot}).
Qualitatively, we obtain the same picture as in the crude estimate
discussed at the beginning (up to factors like $c'$).

The evolution of this crossover regime with impurity concentration is
shown in Fig.~\ref{figRhoLogLog} and its inset. The dependence of the
distribution function (and of $\Delta \rho$) on impurity-concentration
is a reflection of the complete breakdown of Matthiessen`s rule in the
crossover regime, where it is not possible to separate the different
scattering mechanisms contributing to the resistivity. In addition,
while not entirely physically meaningful, the ($T$-dependent)
effective exponent defined by the logarithmic derivative of
$\rho(T)-\rho(0)$ in Fig.~\ref{figExponent}, when properly
interpreted, displays the various crossovers in a dramatic way.  For
example, even for reasonably clean system either asymptotic exponents,
$2$ and $1.5$ are difficult to observe, while effective exponents
close to $1$ dominate over a wide range of parameter values as suggested
by our estimate (\ref{alpha}).

\begin{figure}[t]
  \centering
 \epsfig{width=\linewidth,file=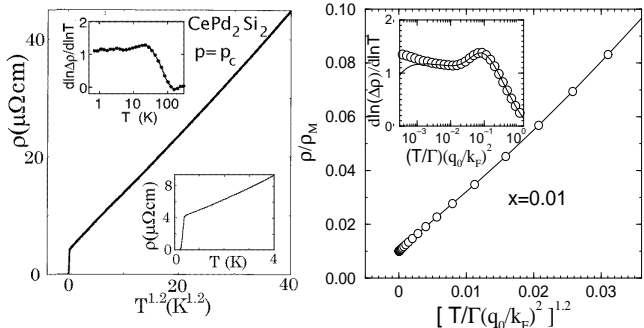}
\caption[]{Resistivity as a function of $T^{1.2}$ 
  in CePd$_2$Si$_2$ taken from \cite{julian} (left figure)
 compared
  to our calculation (right figure) for $x=0.01$. 
  The insets show the corresponding
  logarithmic derivative of $\rho(T)-\rho(0)$. The solid line in the
  inset of the theoretical plot displays the logarithmic derivative of
  $\rho(T)-0.995 \rho(0)$.  Below $\approx 400$mK CePd$_2$Si$_2$ is
  superconducting (lower inset).
  Note the offset of the line $T=0$ in
  both plots.}
\label{figExperiment}
\end{figure}

While the behavior for $1/(\kf \xi)^2, t \gtrsim 1$ are highly
non-universal and strongly affected by details of the band structure
and scattering mechanisms, this is not the case in the opposite limit
$1/(\kf \xi)^2, t \ll 1$, which is of interest here.  In the latter
regime crossover effects depend only very weakly on the precise
details of the model.

We argue that our approach explains the anomalous resistivity at the
QCP observed in the very good samples of CePd$_2$Si$_2$,
CeNi$_2$Ge$_2$ and CeIn$_3$ \cite{julian,mathur}.  
In Fig.~\ref{figExperiment} we show the
measurement by Grosche {\it et al.} \cite{julian} of the
resistivity as a function of $T^{1.2}$ at the critical pressure in
CePd$_2$Si$_2$ and compare them to our model for $x=0.01$. It is
important to note, that not only the same exponent of the resistivity
shows up in the theory, but that it is also observed over a similar
range $0.1 \rho_0 \lesssim \Delta \rho(T) \lesssim 10 \rho_0$. This
suggests that standard spin-fluctuation theory
\cite{hertz,millis,moriya} can be applied for this system opening the
possibility for a theory of the striking superconducting phase
observed below $400$mK.  In CeNi$_2$Ge$_2$, a similar exponent is
observed in the resistivity \cite{julian}, while the effective
exponents in samples of CeIn$_3$ \cite{mathur} show a behavior
reminiscent of our predictions for $x=0.1$ in Fig.~\ref{figExponent}.
It is certainly necessary to check
other predictions of this theory carefully. For example, according
to \cite{millis}, the pressure dependence of the
N\`eel temperature near the QCP should be given by $(p-p_c)^{2/3}$
while experimentally a linear dependence seems to be observed over
some range
\cite{julian}.  Also the specific heat should give valuable
informations on the nature of the spin fluctuations. The most direct
test of the effects described in this letter 
is, however, a comparison of the critical
resistivity in samples of different quality. According to our theory,
the effective exponent has to change from $1.5$ for dirty samples to
values near $1$ for very clean samples.  Also, for cleaner and cleaner
samples, a ``bump'' has to show up in plots of the effective
 exponent (cf. Fig.~\ref{figExponent}).
 The dependence of the exponent on sample quality has
indeed been reported \cite{julian}.

For systems not directly at but still near
the QCP ($\kf \xi(T=0)\gg 1$),
 we expect again a large crossover regime in the
resistivity with anomalous effective exponents due to a 
pronounced crossover from $\rho\approx \rho_M (T/\Gamma)^2$ at high
temperatures to  $\rho\approx \rho_M(x+ (T/\Gamma)^2 (\kf \xi))$
at lowest temperatures.

I would like to thank N.~Andrei, P.~Coleman, M.~Grosche, 
L.~Joffe, G.~Kotliar, H.~v.~L\"ohneysen,
 A.~Millis, C.~Pfleiderer, A.~Ruckenstein and
 P.~W\"olfle for discussions and the
 A.~v.~Humboldt
foundation for financial support.

\end{document}